\documentclass[twocolumn,letterpaper,nofootinbib,prd,amsmath,floatfix]{revtex4}

\usepackage{epsfig}
\usepackage{hyperref}

\newcommand{\un}{{\underline n}}
\newcommand{\unp}{{\underline{n+1}}}
\newcommand{\plusP}{{{}^{\scriptscriptstyle (+)}{\cal P}}}
\newcommand{\minusP}{{{}^{\scriptscriptstyle (-)}{\cal P}}}
\newcommand{\lessthansimilarto}{\lower3pt\hbox{$\buildrel{<}\over{\sim}$}}
\newcommand{\greaterthansimilarto}{\lower3pt\hbox{$\buildrel{>}\over{\sim}$}}

\begin{document}

\title{
The Midpoint Rule as a Variational--Symplectic Integrator. \\
I. Hamiltonian Systems}

\author{J.~David Brown}
\affiliation{Department of Physics, North Carolina State University,
Raleigh, NC 27695 USA}

\begin{abstract}
Numerical algorithms based on variational and symplectic integrators exhibit 
special features that make them promising candidates for application to general 
relativity and other constrained Hamiltonian systems. This paper lays 
part of the foundation for such applications. The midpoint rule for 
Hamilton's equations is examined from the perspectives of variational 
and symplectic integrators. It is shown that the midpoint rule preserves 
the symplectic form, conserves Noether charges, and exhibits excellent 
long--term energy behavior. The energy behavior is explained by the 
result, shown here, that the midpoint rule exactly conserves a phase 
space function that is close to the Hamiltonian. The presentation includes
several examples.
\end{abstract}
 
\maketitle

\section{Introduction}
This is the first in a series of papers that  explore the possible advantages of 
using variational and symplectic numerical integration techniques for constrained 
Hamiltonian systems. A constrained Hamiltonian system is the Hamiltonian formulation 
of a gauge theory \cite{HenneauxTeitelboim}. 
For such a theory  the canonical momenta, defined by the derivatives of the 
Lagrangian with respect to the velocities, are not invertible for the velocities 
as functions of the coordinates and momenta. As Dirac showed \cite{Dirac}, this implies the 
presence of constraints among the coordinates and momenta. The constraints are 
the canonical generators of the gauge 
symmetry. They appear in the action as part of the Hamiltonian, accompanied by
undetermined multipliers. 

Constrained Hamiltonian systems are common in physics. 
Examples include electrodynamics, Yang--Mills theories, 
string theory, and general relativity. The numerical integration of Maxwell's equations 
for electrodynamics has been well studied. For example, with the finite difference time domain (FDTD) 
method, the electric and magnetic fields are evolved using discrete forms of 
Ampere's and Faraday's laws \cite{TafloveHagness}. 
The FDTD discretization automatically preserves the two Gauss's law constraints in the source free 
case. Yang--Mills and string theories are primarily used to describe 
elementary quantum systems, so for these theories the classical solutions do not play a critical 
role. Correspondingly, numerical methods for evolving the classical Yang--Mills fields and classical 
strings have not been thoroughly explored. 

The most challenging example of a constrained Hamiltonian system, and the 
one that serves as my primary motivation for this investigation, is general relativity. 
There is currently a great deal of interest in developing numerical methods for solving 
Einstein's equations. This interest is driven by recent advances on the experimental front. A number 
of ground--based gravitational wave detectors are in operation today, and during the next decade 
some of these instruments will reach the level of sensitivity needed to detect black 
hole collisions. The 
LISA project is a joint effort between NASA and ESA, with the goal of placing a 
gravitational wave detector in solar orbit. The LISA detector will be capable of sensing, among 
other sources, collisions between the supermassive black holes that reside at the centers of  
galaxies. To maximize the scientific payoff of these instruments we need a theoretical 
understanding of the gravitational--wave signals produced by black hole collisions and other 
astrophysical phenomena. The only known method for predicting the gravitational wave signature of 
colliding black holes is through numerical simulation. 

Numerical relativity is not a mature field. Researchers have spend much 
time and effort in developing numerical relativity codes, but the complexity of the Einstein equations 
coupled with the topological issues that arise when modeling black holes have made progress slow. 
Current codes can succeed in simulating at most about one orbit of a binary black hole system 
before errors completely spoil the results \cite{Bruegmann:2003aw}. The main difficulty appears 
to be the presence of 
``constraint violating modes'' \cite{Kidder:2000yq,Kidder:2001tz,Scheel:2002yj,Lindblom:2004gd}. 
These are solutions of the Einstein evolution equations that 
are unphysical in that  they do not respect the constraints. Although the evolution equations 
preserve the constraints at an analytical level, numerical errors inevitably excite these 
constraint violating modes. Some of these modes grow exponentially fast and spoil the 
numerical results. 
What is needed for numerical relativity is an algorithm that will keep the 
constraints satisfied, or nearly satisfied, during the course of the 
evolution. It might be possible to develop a scheme like the FDTD method of 
electrodynamics, but the 
complexity and nonlinearity of the Einstein equations makes this a difficult task. Some 
progress along these lines has been made by Meier \cite{Meier:2003bn}.

In this paper I begin to explore a different route for keeping the constraints satisfied for 
general relativity and other constrained Hamiltonian systems. The idea is based on the 
use of variational integrators (VI). In the traditional approach to numerical modeling 
by finite differences, the continuum equations of motion  are
discretized by replacing derivatives with finite difference approximations. 
In the VI approach we first discretize the action, then derive the 
discrete equations of motion by extremizing the action. This approach was pioneered 
by a number of researchers beginning in the 1960's; for a brief historical overview, 
see Ref.~\cite{Marsden:2001}. Variational integrators have been developed further in recent years 
by Marsden and collaborators \cite{Marsden:2001,Lew:2004}. 

One of the key properties 
of variational integrators is that they are symplectic. This means that the discrete 
time evolution defined by the VI equations automatically conserves a symplectic form. The subject of 
symplectic integrators is well--developed; for an overview, see Ref.~\cite{SanzSerna}. 
Variational integrators also conserve the charges 
associated with symmetries via Noether's theorem. For our present purposes, the most 
interesting characteristic of variational and symplectic integrators is their behavior 
regarding energy. Although these integrators do not typically conserve energy, they exhibit 
excellent long--time energy behavior. For other integrators the energy errors typically 
increase unboundedly in time. For variational and symplectic integrators the energy 
error is typically  bounded in time. 

There are various ways that one can develop a variational integrator for 
constrained Hamiltonian systems. For example, one can extremize the action while keeping the 
undetermined multipliers fixed. In that case   the constraints will not remain 
zero under the discrete time evolution. But there is reason to believe that in many 
cases the constraint errors, like energy, will remain bounded in time \cite{Browninprep}. 
Another option is to extremize the action with respect to the undetermined multipliers 
as well as the canonical coordinates and momenta. This is the most 
attractive approach from a number of perspectives. In this case the discrete constraints are 
imposed as equations of motion at each timestep, so they are guaranteed to hold under the 
discrete time evolution. The trade off is that the undetermined multipliers of the continuum 
theory are actually determined by the discrete equations of motion. 

In general relativity the constraints cannot be solved 
for the multipliers unless the coordinates and momenta are chosen appropriately. 
The traditional choice of canonical coordinates \cite{Arnowitt:1962hi}, the spatial metric, 
leads to generically ill--defined equations for the multipliers. Recently Pfeiffer 
and York \cite{York:1998hy,Pfeiffer:2002iy} have rewritten the constraints using
the conformal metric and the trace of the extrinsic curvature as coordinates. 
They show that the resulting equations for the multipliers are generically well--defined. 
In Ref.~\cite{Brown:2005aq} I rewrote the action and evolution equations in terms of these new 
coordinates and their conjugate momenta. This is one form of the action that is suitable
for the development of a  variational integrator for general relativity. 

The essential idea of using a discrete action to define a set of discrete equations of motion 
that both respect the constraints and determine the multipliers has also been studied in the 
context of general relativity by Di Bartolo, Gambini and Pullin
\cite{DiBartolo:2002fu,Gambini:2002wn,Gambini:2002zi,DiBartolo:2004dn,Gambini:2005za,Gambini:2005sv}.
They refer to their approach as ``consistent discretization''. They  discuss consistent 
discretization 
in the context of numerical relativity, and also as a  route toward 
quantization. There are a number of technical differences between the works of 
Di Bartolo, Gambini and Pullin and the results presented in this and the following papers. 
The most important difference between my approach and theirs is a difference in 
techniques used to generate the equations of motion. I extremize the discrete action 
directly while Di Bartolo {\it et al.} identify  the discrete Lagrangian as the generator 
of a Type 1 canonical transformation. With direct extremization we obtain useful information 
about the system encoded in the endpoints of the varied action. This is the key to 
proving the important properties of the variational integrator including symplecticity, 
Noether's theorem, and the good long--time behavior of energy. 

In this first paper I focus on simple mechanical systems with no constraints. This is a 
rich subject that has been explored rather thoroughly, in mathematically precise 
language, by 
Marsden {\it et al.} \cite{Marsden:2001,Lew:2004}. 
The purpose of this paper is to present  the key results on variational integrators 
in the context of a particular discretization using language familiar to  most 
physicists. The particular discretization of the action considered here leads to the midpoint 
rule applied to Hamilton's equations. The midpoint rule is an old, familiar numerical algorithm.
It is presented here in a new, perhaps unfamiliar light as a variational--symplectic integrator. This 
new perspective allows us to derive and to understand the characteristic features of this integrator 
on a rather deep level. 

In the next section I review the derivation of Hamilton's equations from the action 
expressed in Hamiltonian form. In Sec.~\ref{sectionthree}, I discretize the action and derive 
the VI equations from its extremum. In Sec.~\ref{sectionfour} I show that the variational 
integrator is symplectic, and Noether's theorem holds. I also show that the VI equations 
can be written as the midpoint rule applied to Hamilton's equations. Section \ref{sectionfive} 
contains a discussion of energy. There, it is shown that the energy is well behaved because the 
VI equations exactly conserve the value of a phase space function that is close, in a sense to 
be discussed, to the Hamiltonian. Several examples are given in Sec.~\ref{sectionsix}. These examples
explore the energy behavior and the convergence properties of the midpoint rule as a 
variational integrator. 

My goal is to investigate variational and symplectic integration techniques for constrained 
Hamiltonian systems. In the next paper in this series \cite{Browninprep},  
I will apply these techniques to 
a class of simple constrained Hamiltonian systems, namely,  parametrized 
Hamiltonian mechanics. 
These are ordinary Hamiltonian systems with the coordinates, momenta, and time expressed as 
functions of an arbitrary parameter. The theory is invariant under changes of the parameter, and 
this  gauge invariance gives rise to a constraint that enforces conservation of energy. 
In future papers I will apply  VI techniques to field theories with gauge symmetries. In 
canonical form these theories are described as 
constrained Hamiltonian systems with constraints that are local functions in space. 

\section{Continuum Mechanics}\label{sectiontwo}
Let the index $a$ label pairs of canonically conjugate dynamical variables $x_a$ and $p_a$. 
The action is a functional of $x_a(t)$ and $p_a(t)$, given by 
\begin{equation}\label{S_continuous}
  S[p,x] = \int_{t'}^{t''} dt \left[ p_a \dot{x}_a
    - H(p,x,t) \right] \ .
\end{equation}
Here, $H(p,x,t)$ is the Hamiltonian and $t$ is physical time.
The dot denotes differentiation with respect $t$. The summation convention is used for repeated indices, 
so the expression $p_a \dot{x}_a$ includes an implied sum over $a$. 

Variation of the action (\ref{S_continuous}) yields
\begin{eqnarray}\label{deltaS_continuous}
  \delta S[p,x] & = & \int_{t'}^{t''} dt \left[ \left( \dot{x}_a
    - \frac{\partial H}{\partial p_a} \right)
    \delta p_a  \right.  \nonumber\\
      & & \qquad{\ }\left.  + \left( -\dot{p}_a 
    - \frac{\partial H}{\partial x_a} \right) \delta x_a   \right] \nonumber \\
  & & \qquad{\ } + \, p_a \delta x_a \Bigr|_{t'}^{t''}  \ .
\end{eqnarray}
With the coordinates $x_a$ fixed at the initial and final times, $t'$ and $t''$, the 
endpoint terms in $\delta S$ vanish. 
Then the condition that the action should be stationary, $\delta S = 0$, yields
\begin{subequations}\label{HamiltonsEquations}
  \begin{eqnarray}
    \dot{x}_a  & = & \frac{\partial H}{\partial p_a} \ ,\label{HamiltonsEquationsA}\\
    \dot{p}_a  & = & - \frac{\partial H}{\partial x_a} \ , \label{HamiltonsEquationsB}
  \end{eqnarray}
\end{subequations}
These  are the familiar Hamilton's equations. An immediate consequence of these equations 
is that the Hamiltonian function $H(p,x,t)$ satisfies
\begin{equation}\label{Hdot}
  \dot{H} = \frac{\partial H}{\partial t} \ .
\end{equation}
If $H$ has no explicit $t$ dependence, then $\dot H = 0$. In this case $H$, the energy, is a constant 
of the motion.

\section{Discrete Mechanics}\label{sectionthree}
\begin{figure}[htb]
\includegraphics{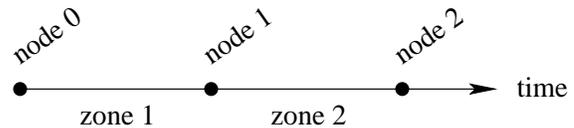}
\caption{Discretization in time. The nodes are labeled $n = 0,\ldots,N$ and the 
zones (or time intervals) are labeled $n = 1,\ldots,N$. The coordinates and 
momenta are node centered, the Hamiltonian function is zone centered.}
\label{TimeAxis.figure}
\end{figure}

Let us divide the time interval between $t'$ and $t''$ into $N$ equal subintervals,  or ``zones'', 
labeled $n=1, \ldots , N$. These zones are separated by nodes, which are labeled $n = 0,\ldots, N$. 
As seen in Fig.~(\ref{TimeAxis.figure}) the endpoints of zone $n$ are nodes $n-1$ and $n$.
The expression $t^n$ denotes the time at the $n$th node. Likewise, $x_a^n$ and $p_a^n$ denote 
the coordinates and momenta at the $n$th node. The timestep is $\Delta t = t^n - t^{n-1}$. 
In this paper I consider the following second order accurate discretization of the action 
(\ref{S_continuous}): 
\begin{equation}\label{S_discrete}
  S[p,x] = \sum_{n=1}^{N} \Delta t \left[ p^\un_a\, \frac{\Delta x^n_a}{\Delta t} -
    H(p^\un,x^\un,t^\un) \right] \ .
\end{equation}
The  $\Delta$ notation and the underlined index notation are employed repeatedly below; they are 
defined by 
\begin{subequations}\label{keydefs}
\begin{eqnarray}
  \Delta x^n_a & \equiv & x^n_a - x^{n-1}_a \ , \label{keydefs1}\\
  x_a^\un  & \equiv & \frac{x_a^n + x_a^{n-1}}{2} \ . \label{keydefs2}
\end{eqnarray}
\end{subequations}
These operations commute; that is, $x_a^\un - x_a^{\underline{n-1}} = (\Delta x_a^n + \Delta x_a^{n-1})/2$. 

It will also prove useful to denote the value of the Hamiltonian in the $n$th  zone by   
\begin{equation}\label{keydefsH}
  H^n \equiv H(p^\un,x^\un,t^\un)
\end{equation}
That is, we view $t$, $x_a$, and $p_a$ as 
node centered in time and $H$ as zone centered in time. [See Fig.~(\ref{TimeAxis.figure}).] 
Then equation (\ref{keydefsH}) expresses 
the fact that, to second order accuracy, the zone centered values of $t$, $x_a$, and $p_a$ that 
appear in $H^n$ can be obtained from the averages of the neighboring node centered values. 

The discrete ``Lagrangian'', that is, the term in square brackets in Eq.~(\ref{S_discrete}), 
has truncation errors that scale 
like ${\cal O}(\Delta t^2)$. The discrete action is a sum over $N \sim 1/\Delta t$ terms, each having 
errors of order ${\cal O}(\Delta t^3)$. It follows that the error in 
$S$ typically scales like ${\cal O}(\Delta t^2)$. 
Thus the action (\ref{S_discrete}) is second order accurate.  
Note that Eq.~(\ref{S_discrete}) is not the only possible second 
order discretization of the action. For pedagogical purposes, I have chosen to restrict considerations in this 
paper to the discrete action (\ref{S_discrete}). Other discretizations, 
including some with higher order accuracy, 
will be discussed elsewhere \cite{Browninprep}.

Note that the momentum variables appear in the action (\ref{S_discrete}) only  in the combination 
$p_a^\un \equiv (p_a^n + p_a^{n-1})/2$. This combination represents 
the zone centered momentum, accurate to second order. Let us set this observation aside for 
the moment and treat the action as a function of all node--centered coordinates and 
momenta, $x_a^n$ and $p_a^n$ for $n = 0,\ldots ,N$. The variation of $S$ is
\begin{eqnarray}\label{deltaS_discrete}
  \delta S & = & \sum_{n=1}^{N-1}\Delta t  \left[ \frac{\Delta x_a^\unp}{\Delta t} -
    \left(\frac{\partial H}{\partial p_a}\right)^\unp 
     \right]   \delta p_a^n   \nonumber  \\
    & + &  \sum_{n=1}^{N-1} \Delta t \left[ -\frac{\Delta p_a^\unp}{\Delta t} - 
    \left(\frac{\partial H}{\partial x_a}  \right)^\unp  \right]\delta x_a^n  
       \nonumber \\ 
    & & + \frac{1}{2} \left[ \Delta x_a^1 - \left( \frac{\partial H}{\partial p_a}\right)^1\Delta t 
       \right] \delta p_a^0   \nonumber \\
     & & + \frac{1}{2} \left[ \Delta x_a^N - \left( \frac{\partial H}{\partial p_a}\right)^N\Delta t 
       \right] \delta p_a^N   \nonumber \\
     & & - \left[ p_a^{\underline{1}} + \frac{1}{2}\left( \frac{\partial H}{\partial x_a} \right)^1 \Delta t 
       \right] \delta x_a^0  \nonumber \\
     & & +  \left[ p_a^{\underline{N}} - \frac{1}{2}\left( \frac{\partial H}{\partial x_a} \right)^N \Delta t 
       \right] \delta x_a^N  \ .
\end{eqnarray}
Here and below we treat the derivatives of $H(p,x,t)$, like $H$ itself, as zone centered 
quantities. Recall the notation defined in Eqs.~(\ref{keydefs}) and (\ref{keydefsH}). For any 
zone--centered function $F(p,x,t)$ of the canonical variables and time, we have 
$F^\unp \equiv [F^{n+1} + F^n]/2 \equiv [F(p^\unp,x^\unp,t^\unp) + F(p^\un,x^\un,t^\un)]/2$. These 
notational rules apply to the derivatives of the Hamiltonian that appear in $\delta S$. 

If we fix the coordinates at the endpoints, $x_a^0$ and $x_a^N$,  then the condition that 
the discrete action should be extremized is
\begin{subequations}\label{discreteequations}
  \begin{eqnarray} 
    \frac{\Delta x_a^\unp}{\Delta t}  & = & \left(\frac{\partial H}{\partial p_a}\right)^\unp 
    \ ,\quad  n = 1,\ldots ,  N{\!}-{\!}1  \ , \label{discreteequationsA}\\
    \frac{\Delta p_a^\unp}{\Delta t} 
    & = & -  \left(\frac{\partial H}{\partial x_a}\right)^\unp 
     \ ,\quad\!  n = 1,\ldots ,  N{\!}-{\!}1 \ ,\qquad  \label{discreteequationsB}\\
    \frac{\Delta x^1}{\Delta t} & = & \left(\frac{\partial H}{\partial p_a}\right)^1 
    \label{discreteequationsC} \\
    \frac{\Delta x^N}{\Delta t} & = & \left(\frac{\partial H}{\partial p_a}\right)^N 
    \label{discreteequationsD}
  \end{eqnarray}
\end{subequations}
These equations  are redundant. For example, Eq.~(\ref{discreteequationsD}) can be derived from 
Eqs.~(\ref{discreteequations}a,c). This redundancy is a result of the fact that the 
action does not depend on the node--centered 
momenta $p_a^n$ independently, but only on the zone--centered combinations $p_a^\un$. 
We can combine equations (\ref{discreteequations}a,c,d) into 
a single expression and write the equations of motion (\ref{discreteequations}) as
\begin{subequations}\label{detwo}
  \begin{eqnarray}
     \frac{\Delta x_a^{n+1}}{\Delta t}  & = & \left(\frac{\partial H}{\partial p_a}\right)^{n+1} 
    \ ,\quad  n = 0,\ldots ,  N{\!}-{\!}1  \ , \label{detwoA}\\
    \frac{\Delta p_a^\unp}{\Delta t} 
    & = & -  \left(\frac{\partial H}{\partial x_a}\right)^\unp 
     \ ,\quad\!  n = 1,\ldots ,  N{\!}-{\!}1 \ ,\qquad  \label{detwoB}
  \end{eqnarray}
\end{subequations}
The equations of motion in this form can be obtained directly from 
the action (\ref{S_discrete}) by extremizing with respect to the node--centered 
coordinates $x_a^n$ and the zone--centered momenta $p_a^\un$.  They are a discrete form 
of Hamilton's equations (\ref{HamiltonsEquations}).

The equations of motion (\ref{detwo}) constitute the variational integrator defined by the 
discrete action (\ref{S_discrete}). Since they are derived from a variational principle, these equations 
naturally define a boundary value problem in which the freely chosen data are divided between 
the endpoints in time. Thus, given the boundary data $x_a^0$ and $x_a^N$, Eqs.~(\ref{detwo}) determine 
the coordinates $x_a^n$ for $n = 1, \ldots, N-1$ and momenta $p_a^\un$ for $n = 1,\ldots, N$. 
We can add boundary terms to the action to change the permitted boundary conditions.
However, in practice, our primary interest is not in 
any of these boundary value problems. Rather, we are interested in solving an initial value problem.
Thus, we are faced with the task of reinterpreting the equations of motion in such a way that 
initial data can be posed and evolved into the future. 

It is not difficult 
to reinterpret the variational integrator (\ref{detwo}) as an initial value problem. 
One possibility is to choose values for the coordinates at the initial time $t^0$ and 
values for the momenta at the half timestep $t^{\underline 1}$; that is, we  
choose $x_a^0$ and $p_a^{\underline 1}$. Then Eq.~(\ref{detwoA}) with $n=0$ can be 
solved for $x_a^{1}$. This completes the determination of data at  ``levels'' $n=0$ and $1$.
Alternatively, we can generate data at levels $n=0$ and $1$ by
specifying $x^0_a$ and $x^1_a$,  then solving Eq.~(\ref{detwoA}) with $n=0$ 
for $p_a^{\underline 1}$. Once the data at levels $0$ and $1$ have been found, we can 
solve  Eqs.~(\ref{detwo}) with $n=1$  for the level $2$ data
$x_a^2$, and $p_a^{\underline 2}$. We continue in this fashion to obtain the data 
at levels $3$, $4$, {\it etc}.

Strictly speaking, neither of the options outlined above is an {\it initial} value problem. With 
the first option, the freely specifiable data $x_a^0$, $p_a^{\underline 1}$ are split between 
the initial time node and the first time zone. With the second option, the data $x_a^0$ and
$x_a^1$ are split between time nodes $0$ and $1$. Apart from this slight misuse of the word 
``initial'', 
we see that it  is fairly trivial to reinterpret the variational integrator Eqs.~(\ref{detwo}) 
as an initial value problem. With higher order discretizations, this reinterpretation is not so 
simple \cite{Browninprep}. 

\section{Symplectic form,  Noether's theorem and the midpoint rule}\label{sectionfour}
In this section we show that the variational integrator (\ref{detwo}) is symplectic, 
and that Noether's theorem applies. 
These results are derived in mathematically precise language for the Lagrangian formulation 
of mechanics by Marsden {\it et al.} 
\cite{Marsden:2001,Lew:2004}. 
In the process of developing these results, we show that the 
VI equations can be expressed in terms of the node--centered momentum. The discrete equations 
are equivalent to the midpoint rule applied to Hamilton's equations. 

Consider first the continuous Hamiltonian system defined by the action (\ref{S_continuous}). 
The canonical two--form is defined by 
\begin{equation}
  \omega = dp_a \wedge dx_a \ ,
\end{equation}
where $d$ is the exterior derivative and $\wedge$ is the exterior product. Hamiltonian systems are
symplectic, meaning that the  form $\omega$ is 
invariant under time evolution. We can derive this result by noting that, for a 
solution of the classical equations of motion (\ref{HamiltonsEquations}), the variation 
of the action reduces to the endpoint terms in Eq~(\ref{deltaS_continuous}). Let 
$S(x'',t'';x',t')$ denote the action evaluated along the classical history with endpoint 
data $x_a(t') = x'_a$ and  $x_a(t'') = x''_a$. 
We see 
that $\partial S(x'',t'';x',t') /\partial x''_a = p_a(t'')$, and 
$\partial S(x'',t'';x',t') /\partial x'_a = -p_a(t')$. Then the exterior 
derivative of the action is given by 
\begin{equation}\label{dScl_continuous}
  dS(x'',t'';x',t') = p_a \, dx_a \Bigr|_{t'}^{t''} \ .
\end{equation}
where $p_a$ is the canonical momentum evaluated along the classical path. The identity 
$dd\,S(x'',t'';x',t') = 0$ shows that $dp_a \wedge dx_a \Bigr|_{t'}^{t''}$ vanishes, 
so that $\omega$ is constant in time. 

Now turn to the discrete system defined by the action (\ref{S_discrete}). Let us define the 
coefficient of $\delta x_a^N$ that appears in $\delta S$ [Eq.~(\ref{deltaS_discrete})] as ${\minusP}_a^N$, 
where 
\begin{equation}\label{Pminus}
  {\minusP}_a^n \equiv  p_a^{\un} - \frac{\Delta t}{2} 
    \left( \frac{\partial H}{\partial x_a}\right)^{n} \ .
\end{equation}
Similarly, we can define the coefficient of $\delta x_a^0$ as ${\plusP}_a^0$, where 
\begin{equation}\label{Pplus}
  {\plusP}_a^n \equiv  p_a^{\unp} + \frac{\Delta t}{2} 
    \left( \frac{\partial H}{\partial x_a}\right)^{n+1} \ .
\end{equation}
The VI equations (\ref{detwo}) define an evolution in phase space. Obviously 
this evolution can be extended to values of $n$ beyond nodes $0$ and $N$. 
Likewise, we can apply our definitions of $\minusP_a^n$ and $\plusP^n_a$ for all integer $n$. 
Now observe that the VI equations imply that $\plusP^n_a - \minusP^n_a$ vanishes 
when the extended equations of motion hold. Thus, we can drop the 
superscripts $(+)$ and $(-)$ and denote both $\plusP^n_a$ and $\minusP^n_a$ by ${\cal P}^n_a$. 

Let $S(x^N,t^N;x^0,t^0)$ denote the value of the discrete action (\ref{S_discrete}) 
for a solution of the VI equations of motion with endpoint 
data $x_a^N$ at $t^N$ and  $x_a^0$ at $t^0$.
The variation in Eq.~(\ref{deltaS_discrete}) shows that, when the (extended) equations 
of motion hold, 
\begin{equation}\label{discrete_dS}
  dS(x^N,t^N;x^0,t^0) =  {\cal P}^n_a dx^n_a\Bigr|_{n=0}^N  \ .
\end{equation}
This is the analog of Eq.~(\ref{dScl_continuous}) above. 
Taking the exterior 
derivative of this expression we find 
\begin{equation}
  0 = d\, {\cal P}_a^n \wedge dx^n_a \Bigr|_{n=0}^{N} \ .
\end{equation}
Thus, the discrete action naturally defines a symplectic two--form 
\begin{equation}\label{VIsymplecticform}
  \omega = d\, {\cal P}^n_a \wedge dx^n_a 
\end{equation}
that is conserved under the phase space evolution defined by the VI equations of motion.

In the analysis above we defined 
    \begin{eqnarray}\label{nodemomentumdefinition}
      {\cal P}^n_a & = & p_a^{\un} - \frac{\Delta t}{2} 
    \left( \frac{\partial H}{\partial x_a}\right)^{n}  \nonumber\\
    & = & p_a^{\unp} + \frac{\Delta t}{2} 
    \left( \frac{\partial H}{\partial x_a}\right)^{n+1} \ .
    \end{eqnarray}
The two expressions for ${\cal P}^n_a$ are equivalent when the equations of 
motion hold. A short calculation using Eq.~(\ref{detwoB}) shows that 
\begin{equation}\label{averagenodemomentum}
  {\cal P}_a^{\unp} \equiv \frac{{\cal P}_a^{n+1} + {\cal P}_a^n}{2} = p_a^{\unp} \ .
\end{equation}
Therefore we see that, when the equations of motion hold, ${\cal P}_a^n$ can be identified 
as the node momentum $p^n_a$. 

The equation of motion for ${\cal P}_a^n$ can be derived by computing 
$\Delta {\cal P}_a^n$ and using the VI equation (\ref{detwoB}). Along with Eq.~(\ref{detwoA}), 
we have 
\begin{subequations}\label{VIeqns}
  \begin{eqnarray}
     \frac{\Delta x_a^{n+1}}{\Delta t}  & = & \left(\frac{\partial H}{\partial p_a}\right)^{n+1} 
     \ , \label{VIeqnsA}\\
    \frac{\Delta {\cal P}_a^{n+1}}{\Delta t} 
    & = & -  \left(\frac{\partial H}{\partial x_a}\right)^{n+1} 
    \ . \label{VIeqnsB}
  \end{eqnarray}
\end{subequations}
This is perhaps the most elegant form of the VI equations. They are simply Hamilton's equations discretized 
with the midpoint rule. Their interpretation as an 
initial value problem is straightforward: given data $x_a^0$ and ${\cal P}_a^0$ at the initial time, 
we solve the equations with $n=0$  for $x_a^1$, ${\cal P}_a^1$. Repeat to find 
data at nodes $n=2, 3, \ldots$.  Recall that the derivatives 
of $H$ that appear on the right--hand 
sides of Eqs.~(\ref{VIeqns}) are zone centered functions. Thus, they are evaluated at ${\cal P}_a^{\unp}$ 
$x_a^{\unp}$, and $t^{\unp}$. 

Noether's theorem states that symmetries give rise to conserved ``charges''. 
We now show that when the discrete action is invariant under a symmetry transformation, 
there exists a charge that is exactly conserved by the VI equations. 

Consider first the continuum case. 
Let $x_a \rightarrow X_a^\sigma(x)$ be a one--parameter family of transformations that leave the 
action, expressed in Lagrangian form, unchanged. Since we are working with the Hamiltonian formalism, 
let us extend this family of configuration space transformations to a family of point canonical 
transformations: 
\begin{subequations}\label{canonicaltransformation}
  \begin{eqnarray}
    x_a & \rightarrow & X_a^\sigma \ ,\\
    p_a & \rightarrow & P_a^\sigma \equiv p_b \frac{\partial X^{-\sigma}_{b}}{\partial x_a} \ .
  \end{eqnarray}
\end{subequations}
Here, it is assumed that $\sigma = 0$ coincides with the identity transformation. By 
differentiating 
the relation $x_a = X_a^{-\sigma}(X^\sigma(x))$ we see that $P_a^\sigma {\dot X}_a^\sigma = p_a {\dot x}_a$ 
so the transformation (\ref{canonicaltransformation}) is indeed canonical. 

By assumption the action (\ref{S_continuous}) is unchanged by the transformations 
(\ref{canonicaltransformation}), so
we have  $S[p,x] = S[P^\sigma,X^\sigma]$ for all $\sigma$. It follows that the 
derivative of $S[P^\sigma,X^\sigma]$ with respect to $\sigma$ must vanish. On the other hand, 
the endpoint terms in the general variation of the action (\ref{deltaS_continuous}) imply that,
if $X^\sigma_a$, $P^\sigma_a$ satisfy the equations of motion when $\sigma=0$, then
\begin{equation}
   \frac{d S[P^\sigma,X^\sigma]}{d\sigma} = 
   \left. p_a \frac{d X^\sigma_a}{d\sigma} \right|^{t''}_{t'} 
\end{equation}
at  $\sigma = 0$. 
Because the left--hand side of this equation vanishes, we see that the charge
\begin{equation}
  Q \equiv p_a \left.\frac{dX^\sigma_a}{d\sigma}\right|_{\sigma = 0}
\end{equation}
is conserved in time for the classical motion of the system. 

Now turn to the discrete case. Let us assume that the discrete action (\ref{S_discrete}) is 
unchanged when the variables $x_a^n$, $p_a^n$ are transformed by Eqs~(\ref{canonicaltransformation}) 
for each value of $n$. Then, as in the continuum case, the derivative of $S[P^\sigma,X^\sigma]$ 
with respect to $\sigma$ vanishes. The general variation of the action (\ref{deltaS_discrete}) implies 
that, if $(X^\sigma_a)^n$, $(P^\sigma_a)^n$ satisfy the equations of motion at $\sigma=0$, then
\begin{equation}
   \frac{d S[P^\sigma,X^\sigma]}{d\sigma} 
   =  \left. {\cal P}^n_a \frac{d (X^\sigma_a)^n}{d\sigma} \right|_{n=0}^N 
\end{equation}
at $\sigma = 0$. Here, I have used the definitions (\ref{nodemomentumdefinition}) for 
${\cal P}^n_a$. 
Since the left--hand side of this relationship vanishes,  we find that the charge
\begin{equation}
  Q \equiv {\cal P}^n_a \left. \frac{d(X^\sigma_a)^n}{d\sigma}\right|_{\sigma = 0}
\end{equation} 
is conserved (independent of $n$) by the VI Eqs.~(\ref{detwo}) or (\ref{VIeqns}). 

\section{Energy conservation}\label{sectionfive}
One of the key characteristics of variational integrators that makes them interesting and 
important is their behavior with respect to energy. If the Hamiltonian has no explicit 
time $t$ dependence then the energy is conserved in the continuum theory; see Eq.~(\ref{Hdot}). 
Variational integrators do not conserve energy exactly, but typically the energy error 
does not grow as the evolution time increases. To be precise, in this section I show 
that the  VI equations (\ref{VIeqns}) exactly conserve the value of a phase space function ${\cal H}$ 
that differs from the Hamiltonian $H$ by terms of order ${\cal O}(\Delta t^2)$. The coefficient of the 
${\cal O}(\Delta t^2)$ difference is a phase space function that remains bounded at least 
as long as the solution trajectory is bounded in phase space. It follows that  the value 
of energy predicted by the VI equations will be ``close'' to the exact value, where ``close'' 
means that the error is of order ${\cal O}(\Delta t^2)$  with a coefficient that does not 
exhibit unbounded growth in time. 

Let us begin the analysis by considering the continuum evolution for a system with time--independent 
Hamiltonian ${\cal H}(p,x)$. This system is described by the action 
\begin{equation}\label{continuousS}
  {\cal S}[p,x] = \int_{t'}^{t''} dt \left[ p_a {\dot x}_a - {\cal H}(p,x) \right] \ .
\end{equation}
with variation 
\begin{equation}\label{deltaSwithdeltat}
  \delta{\cal S} = {\hbox{eom's}} +  p_a \delta x_a 
    \Bigr|_{t'}^{t''} \ .
\end{equation}
The terms listed as ``eom's'' are the terms that yield Hamilton's equations of motion. 
Let ${\cal S}(x'',t'';x',t')$ denote the action (\ref{continuousS}) evaluated at the solution of 
Hamilton's equations with endpoint data $x'_a$ at $t'$ and $x''_a$ at $t''$. 
The variation Eq.~(\ref{deltaSwithdeltat}) implies that ${\cal S}(x'',t'';x',t')$ satisfies 
\begin{subequations}\label{genfuneqns}
  \begin{eqnarray}
    \frac{\partial {\cal S}(x'',t'';x',t')}{\partial x''_a} & = & p'' \ ,\\
    \frac{\partial {\cal S}(x'',t'';x',t')}{\partial x'_a} & = & -p'  \ ,
  \end{eqnarray}
\end{subequations}
where $p'_a = p_a(t')$ and $p''_a = p_a(t'')$. 
These equations show that $-{\cal S}(x'',t'';x',t')$ is a Type 1 generating function for a canonical 
transformation from ``old'' coordinates and momenta $x'_a$, $p'_a$ to ``new'' coordinates 
and momenta $x''_a$, $p''_a$ \cite{Goldstein}. We also know that, starting from the 
initial data $x'_a$, $p'_a$, 
the classical trajectory generated by the Hamiltonian ${\cal H}(p,x)$ passes through the phase 
space point $x''_a$, $p''_a$. Thus,  $-{\cal S}(x'',t'';x',t')$ is a Type 1 generating function that 
generates a canonical transformation representing the time evolution of the system from
$t'$ to $t''$. 

Generating functions of different type are related by functions of the old and new coordinates 
and momenta. We can define a new generating function $H$ by 
\begin{equation}\label{newgenfxn}
  H  \equiv \frac{p''_a + p'_a}{2} \frac{x''_a - x'_a}{t'' - t'} 
  - \frac{{\cal S}(x'',t'';x',t')}{t'' - t'}  \ .
\end{equation}
Equations (\ref{genfuneqns}) can be written as 
$d{\cal S}(x'',t'';x',t') = p''_a dx''_a 
- p'_a dx'_a$. From this result it is  straightforward to show that the exterior derivative of $H$ 
is given by 
\begin{equation}
  dH = \frac{\Delta x_a}{\Delta t} d{\bar p}_a
  - \frac{\Delta p_a}{\Delta t} d{\bar x}_a  \ ,
\end{equation}
where  ${\bar p}_a \equiv (p''_a + p'_a)/2$, 
${\bar x}_a \equiv (x''_a + x'_a)/2$, $\Delta p_a \equiv p''_a - p'_a$,  
$\Delta x_a \equiv x''_a - x'_a$, and $\Delta t \equiv t'' - t'$. 
Thus, $H$ can be viewed as a function of ${\bar p}_a$ and ${\bar x}_a$.
The canonical transformation that represents the classical evolution from $t'$ to $t''$ 
is written in terms of the new generating function $H \equiv H(\bar p, \bar x)$ as 
\begin{subequations}
  \begin{eqnarray}
    \frac{\partial H({\bar p},{ \bar x})}{\partial {\bar p_a}}  & = & \frac{\Delta x_a}{\Delta t}  \ ,\\
     \frac{\partial H({\bar p}, {\bar x})}{\partial {\bar x_a}}  & = & -\frac{\Delta p_a}{\Delta t}  \ .
  \end{eqnarray}
\end{subequations}
These are precisely the VI equations (\ref{VIeqns}), the midpoint rule,
with some simple changes of notation. 

The analysis above shows that the VI equations (\ref{VIeqns}) can be viewed as the generating 
function equations for a canonical transformation from old coordinates and momenta $x_a^n$, 
${\cal P}_a^n$ to new coordinates and momenta $x_a^{n+1}$, ${\cal P}_a^{n+1}$. 
The generating function is $H({\cal P}^\unp,x^\unp,\Delta t)$; it is helpful  at this point 
in the analysis 
to consider $H$ as dependent on the timestep $\Delta t$ as well as the coordinates and momenta. 
The canonical transformation generated by $H$ defines a mapping of phase space that coincides with the 
exact time evolution described by the Hamiltonian ${\cal H}({\cal P},x)$. 
The relationship between $H$ and ${\cal H}$ is given by 
\begin{eqnarray}\label{HfromcalH}
  H({\cal P}^\unp,x^\unp,\Delta t) & = & {\cal P}^\unp_a \frac{\Delta x_a^{n+1}}{\Delta t}  \nonumber \\
   & & - \frac{{\cal S}(x^{n+1},t^{n+1};x^n,t^n)}{\Delta t} \ .{\quad}
\end{eqnarray}
This is Eq.~(\ref{newgenfxn}) with appropriate changes in notation. 
The function ${\cal S}(x^{n+1},t^{n+1};x^n,t^n)$ is the continuum action (\ref{continuousS}) evaluated at the 
solution of the equations of motion with endpoint data $x_a^n$ at $t^n$ and 
$x_a^{n+1}$ at $t^{n+1}$. Analogous to Eqs.~(\ref{genfuneqns}), we have the relations
\begin{subequations}\label{Pdefns}
  \begin{eqnarray}
    {\cal P}_a^{n+1} & = & \frac{\partial {\cal S}(x^{n+1},t^{n+1};x^n,t^n)}{\partial x_a^{n+1}} \ ,\\
    {\cal P}_a^{n} & = & -\frac{\partial {\cal S}(x^{n+1},t^{n+1};x^n,t^n)}{\partial x_a^{n}} \ .
  \end{eqnarray}
\end{subequations}
that define the momenta at the endpoints. 

The discrete evolution defined by the VI equations with Hamiltonian $H$ coincides
with the exact continuum evolution defined by Hamilton's equations with Hamiltonian 
${\cal H}$. Since the exact evolution conserves ${\cal H}$, it follows that the VI 
equations conserve ${\cal H}$. We now show by explicit calculation that $H$ and ${\cal H}$ 
differ by terms of order ${\cal O}(\Delta t^2)$. 

In order to evaluate the action ${\cal S}$ along the classical solution between $t^n$ and $t^{n+1}$, 
we first expand the solution $x_a(t)$, $p_a(t)$ in a  series in $t$ with coefficients 
that depend on $x_a^\unp$ and $p_a^\unp \equiv {\cal P}_a^\unp$. The calculation is simplified by 
writing Hamilton's equations (\ref{HamiltonsEquations}) as 
\begin{equation}
  {\dot\xi}_a = \omega_{ab}{\cal H}_b \ ,
\end{equation}
where $\xi_a$ denotes the set of canonical variables $\{x_a,p_a\}$ and $\omega_{ab}$ is the matrix 
\begin{equation}
  \omega_{ab} = \left( \begin{array}{cc} 
    0 & -I \\
    I & 0 \end{array} \right) \ .
\end{equation}
Here and below, subscripts on ${\cal H}$ denote derivatives; for example, 
${\cal H}_b \equiv \partial{\cal H}/\partial \xi_b$. The solution is
\begin{eqnarray}\label{xisolution}
 & & \xi_a(t) = \xi_a^\unp \nonumber \\
& & + \omega_{aa'} \left[ {\cal H}_{a'} +  {\cal H}_{a'bc}
  \omega_{bb'}{\cal H}_{b'}\omega_{cc'}{\cal H}_{c'} \Delta t^2/8 \right] (t - t^\unp) \nonumber\\
    & & + \frac{1}{2} \omega_{aa'} {\cal H}_{a'b}\omega_{bb'}{\cal H}_{b'} 
       \left[ (t - t^\unp)^2 - (\Delta t^{n+1})^2/4 \right] \nonumber \\
    & & + \frac{1}{24} \omega_{aa'}  
       \left[ {\cal H}_{a'bc} \omega_{bb'}{\cal H}_{b'}\omega_{cc'}{\cal H}_{c'} 
	 + {\cal H}_{a'b}\omega_{bb'}{\cal H}_{b'c}\omega_{cc'} {\cal H}_{c'}\right] \nonumber \\
 & & \quad \cdot\left[ 4(t - t^\unp)^3 - 3(t - t^\unp)(\Delta t^{n+1})^2\right]  \nonumber \\
       & & + {\cal O}(\Delta t^4) \ ,
\end{eqnarray}
where all derivatives of ${\cal H}$ are evaluated at $\xi_a^\unp$. 

The Type I generating function $-{\cal S}(x^{n+1},t^{n+1};x^n,t^n)$ is written as a function of $\xi_a^\unp$ 
by inserting the solution (\ref{xisolution}) into the action (\ref{continuousS}), with initial and final 
times  $t^n$ and $t^{n+1}$.  The new generating 
function $H$ is then found from Eq.~(\ref{HfromcalH}), with the result 
\begin{equation}\label{HfunctionofcalH}
  H  =  {\cal H} + \frac{1}{24} {\cal H}_{ab}\omega_{aa'}{\cal H}_{a'} 
  \omega_{bb'} {\cal H}_{b'} \Delta t^2 
   + {\cal O}(\Delta t^4) \ .
\end{equation}
Clearly, this formal expansion for $H$ in terms 
of ${\cal H}$ can be inverted to yield 
\begin{equation}\label{calHfunctionofH}
  {\cal H}  =  {H} - \frac{1}{24} {H}_{ab}\omega_{aa'}{H}_{a'} 
  \omega_{bb'} {H}_{b'} \Delta t^2 
   + {\cal O}(\Delta t^4) \ .
\end{equation}
This is the desired relationship between the phase space functions ${\cal H}$ and $H$. 

With the solution $\xi_a(t)$ expanded to terms of order $\Delta t^3$ in Eq.~(\ref{xisolution}), 
the evaluation of Eq.~(\ref{HfromcalH}) yields $H$ through terms of order $\Delta t^2$. 
However, a simple argument can be given to show that the terms of order $\Delta t^3$, and in fact 
all terms proportional to odd powers of $\Delta t$, must vanish. Consider Eq.~(\ref{HfromcalH}), 
but let  the data $t^n$, $x_a^n$ and  $t^{n+1}$, $x_a^{n+1}$ exchange roles. The data must be 
exchanged in the definitions (\ref{Pdefns}) as well; this yields 
\begin{subequations}\label{Pdefns_X}
  \begin{eqnarray}
    {\cal P}_a^{n} & = & \frac{\partial {\cal S}(x^{n},t^{n};x^{n+1},t^{n+1})}{\partial x_a^{n}} \ ,\\
    {\cal P}_a^{n+1} & = & -\frac{\partial {\cal S}(x^{n},t^{n};x^{n+1},t^{n+1})}{\partial x_a^{n+1}} \ .
  \end{eqnarray}
\end{subequations}
Now, the function ${\cal S}(x^{n},t^{n};x^{n+1},t^{n+1})$ is just the action evaluated at the 
solution of Hamilton's equations with endpoint data $x_a(t^n) = x^n_a$ and $x_a(t^{n+1}) = x_a^{n+1}$. 
It differs from ${\cal S}(x^{n+1},t^{n+1};x^{n},t^{n})$ only because the limits of integration are 
reversed. Hence, we have 
\begin{equation}
  {\cal S}(x^{n},t^{n};x^{n+1},t^{n+1}) = - {\cal S}(x^{n+1},t^{n+1};x^{n},t^{n}) \ ,
\end{equation}
and we find that the definitions (\ref{Pdefns_X}) are identical to Eqs.~(\ref{Pdefns}). It follows that the 
right--hand side of Eq.~(\ref{HfromcalH}) is unchanged when we exchange the endpoint data. Equating the 
left--hand sides leads to 
\begin{equation}
  H({\cal P}^\unp,x^\unp,\Delta t) = H({\cal P}^\unp,x^\unp,-\Delta t) \ .
\end{equation}
Therefore, $H$ is an even function of $\Delta t$, and it's expansion (\ref{HfunctionofcalH}) 
in terms of ${\cal H}$ does not contain odd powers of $\Delta t$. 

To summarize, the VI equations exactly conserve ${\cal H}$ and the Hamiltonian $H$ differs 
from ${\cal H}$ by terms of 
order $\Delta t^2$.  The coefficient of the ${\cal O}(\Delta t^2)$ and higher order terms are constructed 
from derivatives of $H$. As long as the motion in 
phase space remains bounded, and the Hamiltonian and its derivatives are nonsingular, then
these coefficients will remain bounded. It follows that $H$ will  
remain ``close'' to ${\cal H}$, which is constant, for all time. 
If the motion in phase space does not remain bounded, it does not necessarily follow that the 
coefficient of the ${\cal O}(\Delta t^2)$ will grow in time. In this situation the results depend 
on the details of the Hamiltonian. 

The energy behavior of the VI equations 
is quite different from the behavior exhibited by  many numerical integrators. For 
example, second--order Runge--Kutta (RK2) typically exhibits errors in $H$ of order $\Delta t^2$ on short 
time scales and a drift in the value of $H$ of order $\Delta t^3$ on long time scales. Fourth 
order Runge--Kutta (RK4) exhibits errors in $H$ of order $\Delta t^4$ on short 
time scales and a drift in $H$ of order $\Delta t^5$ on long time scales. For both RK2 and RK4, 
the energy error becomes unboundedly large as time increases. 
We will see examples of these behaviors in the next section. 

\section{Examples}\label{sectionsix}
The examples in this section show the results obtained from numerical integration of simple 
Hamiltonian systems using the VI equations (\ref{VIeqns}) and standard second and fourth order 
Runge--Kutta. Issues of efficiency are ignored. Clearly the midpoint rule, being implicit, 
is  numerically more 
expensive to solve than explicit integration schemes. However, the aim of this paper is to 
investigate the properties of variational--symplectic integrators without concern for details of
implementation. This is because, ultimately, we would like to apply these methods to theories like 
general relativity  for which standard integration techniques are inadequate. The main goal, 
then, is to find a numerical algorithm that works---efficiency is at most a secondary concern. 

One can  solve the implicit VI equations using  a Newton--Raphson method. But in 
practice it is much simpler and more reliable to iterate the equations until the answer is 
unchanged to a prescribed level of accuracy. Thus, given $x_a^n$ and ${\cal P}_a^n$, we begin the 
first iteration with the approximation $x_a^{n+1} \approx x_a^n$, 
${\cal P}_a^{n+1} \approx {\cal P}_a^n$. This is inserted into the right--hand sides of the 
VI equations to yield  improved approximations for  $x_a^{n+1}$ and ${\cal P}_a^{n+1}$. The whole 
process is repeated until the desired level of accuracy is achieved. 
For the higher resolution runs presented below, about $5$ iterations 
were needed to reach a solution that was accurate to  $1$ part in $10^{13}$. For the lower 
resolution runs, around $15$ iterations were needed to reach this same level of accuracy. 

\subsection{Coupled harmonic oscillators}
\begin{figure}[!htb]
\begingroup%
  \makeatletter%
  \newcommand{\GNUPLOTspecial}{%
    \@sanitize\catcode`\%=14\relax\special}%
  \setlength{\unitlength}{0.1bp}%
\begin{picture}(2448,1943)(0,0)%
\special{psfile=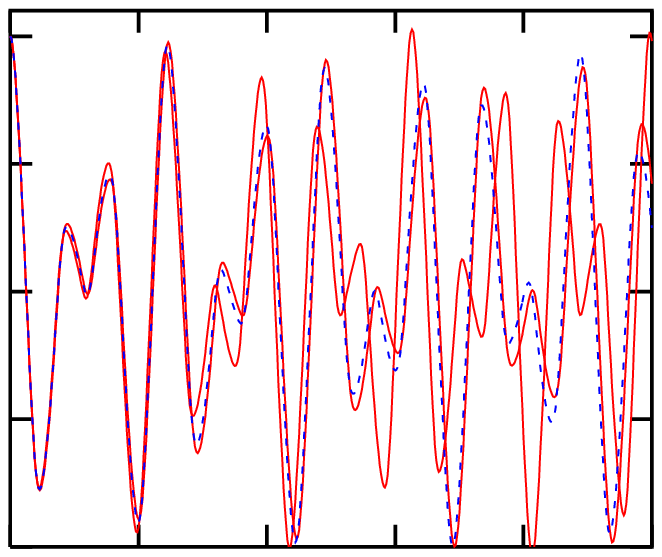 llx=0 lly=0 urx=245 ury=194 rwi=2450}
\put(1660,1660){\makebox(0,0)[l]{${}_{{}_{\displaystyle{\swarrow}}}{\!}$VI}}%
\put(1310,1735){\makebox(0,0)[l]{RK2${}_{{}_{\displaystyle{\searrow}}}$}}%
\put(1374,50){\makebox(0,0){time}}%
\put(100,1072){%
\special{ps: gsave currentpoint currentpoint translate
270 rotate neg exch neg exch translate}%
\makebox(0,0)[b]{\shortstack{oscillator amplitude $x$}}%
\special{ps: currentpoint grestore moveto}%
}%
\put(2298,200){\makebox(0,0){$50$}}%
\put(1928,200){\makebox(0,0){$40$}}%
\put(1559,200){\makebox(0,0){$30$}}%
\put(1189,200){\makebox(0,0){$20$}}%
\put(820,200){\makebox(0,0){$10$}}%
\put(450,200){\makebox(0,0){$0$}}%
\put(400,1770){\makebox(0,0)[r]{$1$}}%
\put(400,1403){\makebox(0,0)[r]{$0.5$}}%
\put(400,1035){\makebox(0,0)[r]{$0$}}%
\put(400,668){\makebox(0,0)[r]{$-0.5$}}%
\put(400,300){\makebox(0,0)[r]{$-1$}}%
\end{picture}%
\endgroup
 
\caption{The amplitude $x$ for the coupled harmonic oscillator as a function of time. 
The VI simulation produces the solid curve that tracks the 
``exact'' solution (dashed curve) fairly closely. The other solid curve is obtained from RK2.}
\label{osc_x}
\end{figure}

Our first example is the system of coupled harmonic oscillators with Hamiltonian 
\begin{equation}
  H = \frac{1}{2}\left(p_x^2 + p_y^2\right) + \frac{1}{2}\left(x^2 + y^2\right) + \frac{1}{5}(x - y)^6 \ .
\end{equation}
The graph in Fig.~\ref{osc_x} shows the amplitude of one of the oscillators, $x$, as a function of 
time. The behavior exhibited is rather complicated. 
The two solid curves show the results of numerical integration with the VI equations 
(\ref{VIeqns}) and second order Runge--Kutta (RK2), both using a timestep of $\Delta t = 0.1$.
The  dashed curve is 
obtained from a fourth order Runge--Kutta integrator with timestep $\Delta t = 0.01$. 
Over the short time scale ($t\leq 50$)  shown in the figure, the dashed curve can be taken 
as the ``exact'' solution. 
Compared to RK2, VI does a visibly better job of tracking the solution.

\begin{figure}[!htb]
\begingroup%
  \makeatletter%
  \newcommand{\GNUPLOTspecial}{%
    \@sanitize\catcode`\%=14\relax\special}%
  \setlength{\unitlength}{0.1bp}%
\begin{picture}(2448,1943)(0,0)%
\special{psfile=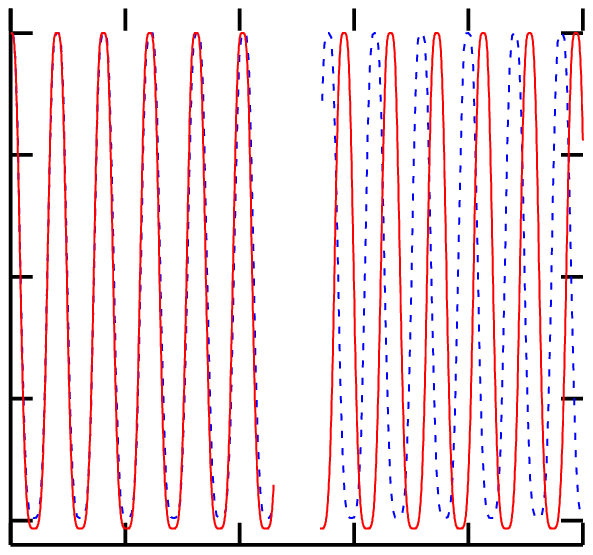 llx=0 lly=0 urx=245 ury=194 rwi=2450}
\put(1450,300){\makebox(0,0)[l]{$\wr\wr$}}%
\put(1474,50){\makebox(0,0){time}}%
\put(100,1072){%
\special{ps: gsave currentpoint currentpoint translate
270 rotate neg exch neg exch translate}%
\makebox(0,0)[b]{\shortstack{energy error (VI)}}%
\special{ps: currentpoint grestore moveto}%
}%
\put(2298,200){\makebox(0,0){$1{\!}\times{\!} 10^6$}}%
\put(1968,200){\makebox(0,0){}}%
\put(1639,200){\makebox(0,0){}}%
\put(1309,200){\makebox(0,0){$10$}}%
\put(980,200){\makebox(0,0){$5$}}%
\put(650,200){\makebox(0,0){$0$}}%
\put(600,1774){\makebox(0,0)[r]{$0$}}%
\put(600,1423){\makebox(0,0)[r]{$-5{\!}\times{\!} 10^{-6}$}}%
\put(600,1072){\makebox(0,0)[r]{$-1{\!}\times{\!} 10^{-5}$}}%
\put(600,721){\makebox(0,0)[r]{$-1.5{\!}\times{\!} 10^{-5}$}}%
\put(600,370){\makebox(0,0)[r]{$-2{\!}\times{\!} 10^{-5}$}}%
\end{picture}%
\endgroup
 
\caption{Energy error for the coupled harmonic oscillator for VI. The solid curve has 
timestep $\Delta t = 0.01$. The dashed curve shows the error divided by $100$ for 
$\Delta t = 0.1$. The results are displayed for the first and last $\sim 12$ time units; 
the total run time was $t = 1\times 10^6$.}
\label{osc_vi_energy}
\end{figure}

The initial data chosen for the coupled oscillator is $x=1$, $y = p_x = p_y = 0$, so the exact 
solution has energy $H = 0.7$. Figure \ref{osc_vi_energy} shows the error in energy 
for  VI at two resolutions. The solid curve shows the error for $\Delta t = 0.01$
while  the dashed curve shows the error divided by $100$ for $\Delta t = 0.1$. 
The close agreement between the amplitudes of these two curves shows that the energy error is second 
order in the timestep. The key observation is that the energy error does not grow in 
time, even for the simulation with a relatively low resolution of $\Delta t = 0.1$.

\begin{figure}[!htb]
\begingroup%
  \makeatletter%
  \newcommand{\GNUPLOTspecial}{%
    \@sanitize\catcode`\%=14\relax\special}%
  \setlength{\unitlength}{0.1bp}%
\begin{picture}(2448,1943)(0,0)%
\special{psfile=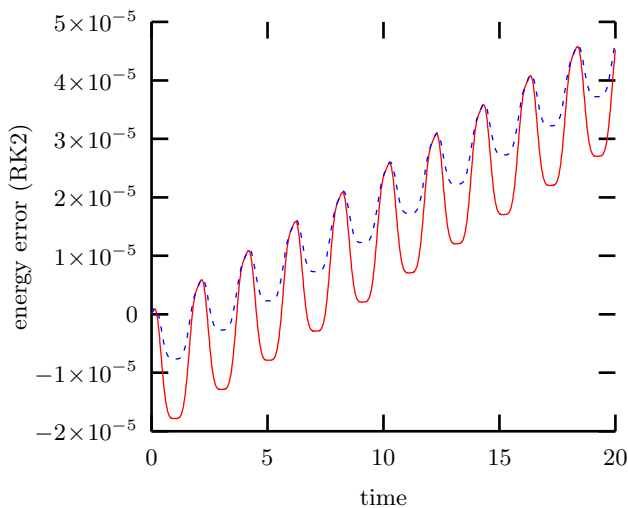 llx=0 lly=0 urx=245 ury=194 rwi=2450}
\put(1424,50){\makebox(0,0){time}}%
\put(100,1072){%
\special{ps: gsave currentpoint currentpoint translate
270 rotate neg exch neg exch translate}%
\makebox(0,0)[b]{\shortstack{energy error (RK2)}}%
\special{ps: currentpoint grestore moveto}%
}%
\put(2298,200){\makebox(0,0){$20$}}%
\put(1861,200){\makebox(0,0){$15$}}%
\put(1424,200){\makebox(0,0){$10$}}%
\put(987,200){\makebox(0,0){$5$}}%
\put(550,200){\makebox(0,0){$0$}}%
\put(500,1844){\makebox(0,0)[r]{ $5{\!}\times{\!} 10^{-5}$}}%
\put(500,1623){\makebox(0,0)[r]{$4{\!}\times{\!} 10^{-5}$}}%
\put(500,1403){\makebox(0,0)[r]{$3{\!}\times{\!} 10^{-5}$}}%
\put(500,1182){\makebox(0,0)[r]{$2{\!}\times{\!} 10^{-5}$}}%
\put(500,962){\makebox(0,0)[r]{$1{\!}\times{\!} 10^{-5}$}}%
\put(500,741){\makebox(0,0)[r]{$0$}}%
\put(500,521){\makebox(0,0)[r]{$-1{\!}\times{\!} 10^{-5}$}}%
\put(500,300){\makebox(0,0)[r]{$-2{\!}\times{\!} 10^{-5}$}}%
\end{picture}%
\endgroup
 
\caption{Energy error for the coupled harmonic oscillator for RK2. The solid curve has timestep 
$\Delta t= 0.01$. The dashed curve shows the error divided by $8$ for $\Delta t = 0.02$.}
\label{osc_rk2_energy}
\end{figure}

The solid curve in Fig.~\ref{osc_rk2_energy} shows the energy error for RK2 at a 
resolution of $\Delta t = 0.01$.  
The dashed curve  is the energy error for RK2 
with resolution $\Delta t = 0.02$, divided by $8$. Note that the two curves 
in this figure coincide on long time scales ($t \agt 5$). This shows that the drift in energy is 
order $\Delta t^3$. The short time scale errors are ${\cal O}(\Delta t^2)$, so the 
``wiggles'' in the low resolution simulation (having been divided by $8$) are 
approximately half the size of the wiggles seen in the high resolution run. 
For this particular system, and this particular choice of initial data, the growth rate 
of the energy error with RK2 is about  $2.5\Delta t^3$ energy units per time unit. 

\begin{figure}[!htb]
\begingroup%
  \makeatletter%
  \newcommand{\GNUPLOTspecial}{%
    \@sanitize\catcode`\%=14\relax\special}%
  \setlength{\unitlength}{0.1bp}%
\begin{picture}(2448,1943)(0,0)%
\special{psfile=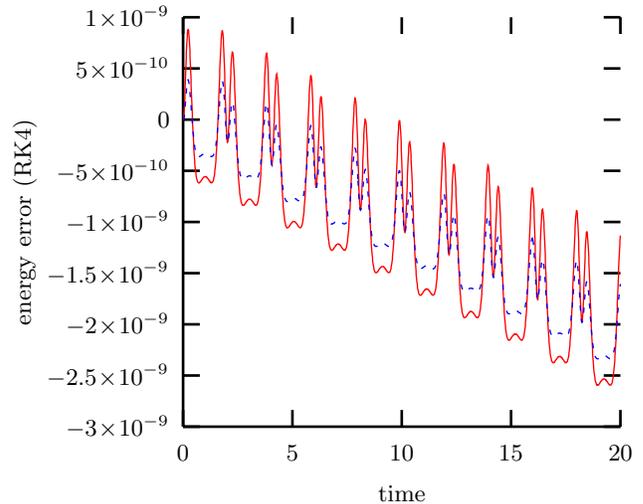 llx=0 lly=0 urx=245 ury=194 rwi=2450}
\put(1474,50){\makebox(0,0){time}}%
\put(100,1072){%
\special{ps: gsave currentpoint currentpoint translate
270 rotate neg exch neg exch translate}%
\makebox(0,0)[b]{\shortstack{energy error (RK4)}}%
\special{ps: currentpoint grestore moveto}%
}%
\put(2298,200){\makebox(0,0){$20$}}%
\put(1886,200){\makebox(0,0){$15$}}%
\put(1474,200){\makebox(0,0){$10$}}%
\put(1062,200){\makebox(0,0){$5$}}%
\put(650,200){\makebox(0,0){$0$}}%
\put(600,1844){\makebox(0,0)[r]{$1{\!}\times{\!} 10^{-9}$}}%
\put(600,1651){\makebox(0,0)[r]{$5{\!}\times{\!} 10^{-10}$}}%
\put(600,1458){\makebox(0,0)[r]{$0$}}%
\put(600,1265){\makebox(0,0)[r]{$-5{\!}\times{\!} 10^{-10}$}}%
\put(600,1072){\makebox(0,0)[r]{$-1{\!}\times{\!} 10^{-9}$}}%
\put(600,879){\makebox(0,0)[r]{$-1.5{\!}\times{\!} 10^{-9}$}}%
\put(600,686){\makebox(0,0)[r]{$-2{\!}\times{\!} 10^{-9}$}}%
\put(600,493){\makebox(0,0)[r]{$-2.5{\!}\times{\!} 10^{-9}$}}%
\put(600,300){\makebox(0,0)[r]{$-3{\!}\times{\!} 10^{-9}$}}%
\end{picture}%
\endgroup
 
\caption{Energy error for the coupled harmonic oscillator for RK4. The solid curve has timestep 
$\Delta t= 0.01$. The dashed curve shows the error divided by $32$ for $\Delta t = 0.02$.}
\label{osc_rk4_energy}
\end{figure}

Qualitatively similar results are found for RK4. In Fig.~\ref{osc_rk4_energy} the solid and dashed 
curves are obtained from simulations with timesteps $\Delta t = 0.01$ and $0.02$,
respectively. 
The errors for the low resolution case have been divided by $32$. 
We see that the long time scale drift in energy is ${\cal O}(\Delta t^5)$, while 
the short time scale ``wiggles'' are ${\cal O}(\Delta t^4)$. 
For this simulation the growth rate of the energy error is about  
$-1.1\Delta t^5$ energy units per time unit. 

The value of energy $H$ obtained from  VI is nearly constant because the VI equations 
exactly conserve the nearby Hamiltonian ${\cal H}$. This can be confirmed by computing the 
first two terms in the expansion for ${\cal H}$ given in Eq.~(\ref{calHfunctionofH}). 
For the coupled harmonic oscillator with timestep $\Delta t = 0.01$, the two--term 
approximation for ${\cal H}$  remains 
nearly constant with variations at the level of $10^{-9}$. With timestep  $\Delta t = 0.1$, 
the approximation for ${\cal H}$  remains nearly constant with variations at the level of $10^{-5}$.
These variations are just what we  expect given the fact that, according to Eq.~(\ref{calHfunctionofH}), 
the terms omitted in the approximation for ${\cal H}$ are order ${\cal O}(\Delta t^4)$. 

\subsection{Simple pendulum}
\begin{figure*}[!htb]
\begingroup%
  \makeatletter%
  \newcommand{\GNUPLOTspecial}{%
    \@sanitize\catcode`\%=14\relax\special}%
  \setlength{\unitlength}{0.1bp}%
\begin{picture}(3600,2160)(0,0)%
\special{psfile=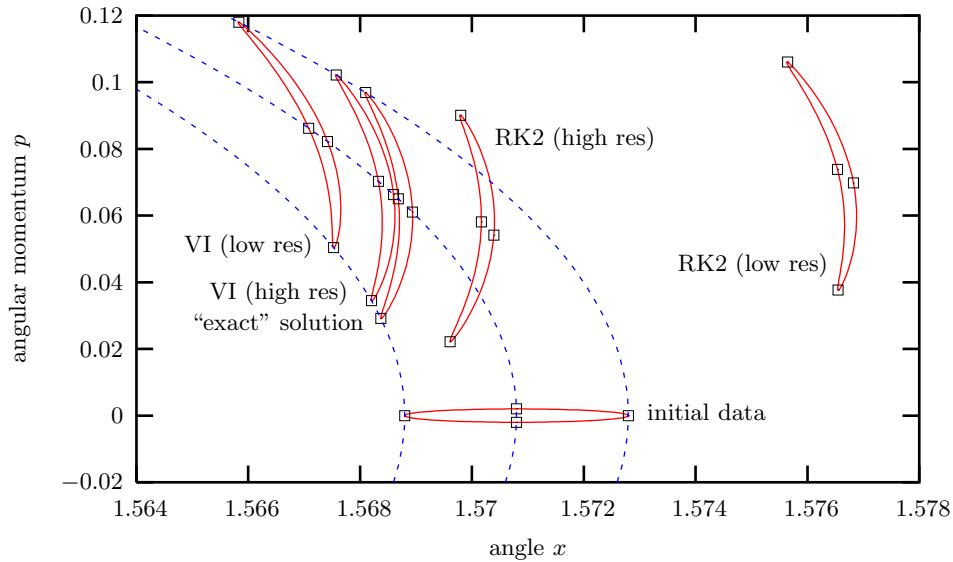 llx=0 lly=0 urx=360 ury=216 rwi=3600}
\put(680,1190){\makebox(0,0)[l]{VI (low res)}}%
\put(775,1015){\makebox(0,0)[l]{VI (high res)}}%
\put(705,905){\makebox(0,0)[l]{``exact" solution}}%
\put(2425,565){\makebox(0,0)[l]{initial data}}%
\put(1850,1595){\makebox(0,0)[l]{RK2 (high res)}}%
\put(2540,1120){\makebox(0,0)[l]{RK2 (low res)}}%
\put(1975,50){\makebox(0,0){angle $x$}}%
\put(100,1180){%
\special{ps: gsave currentpoint currentpoint translate
270 rotate neg exch neg exch translate}%
\makebox(0,0)[b]{\shortstack{angular momentum $p$}}%
\special{ps: currentpoint grestore moveto}%
}%
\put(3450,200){\makebox(0,0){ $1.578$}}%
\put(3029,200){\makebox(0,0){ $1.576$}}%
\put(2607,200){\makebox(0,0){ $1.574$}}%
\put(2186,200){\makebox(0,0){ $1.572$}}%
\put(1764,200){\makebox(0,0){ $1.57$}}%
\put(1343,200){\makebox(0,0){ $1.568$}}%
\put(921,200){\makebox(0,0){ $1.566$}}%
\put(500,200){\makebox(0,0){ $1.564$}}%
\put(450,2060){\makebox(0,0)[r]{ $0.12$}}%
\put(450,1809){\makebox(0,0)[r]{ $0.1$}}%
\put(450,1557){\makebox(0,0)[r]{ $0.08$}}%
\put(450,1306){\makebox(0,0)[r]{ $0.06$}}%
\put(450,1054){\makebox(0,0)[r]{ $0.04$}}%
\put(450,803){\makebox(0,0)[r]{ $0.02$}}%
\put(450,551){\makebox(0,0)[r]{ $0$}}%
\put(450,300){\makebox(0,0)[r]{$-0.02$}}%
\end{picture}%
\endgroup
 
\caption{Phase space diagram for the pendulum. The initial data occupy  
a ``circle'' around $x = \pi/2$, $p = 0$, and are evolved for just under 
$10$ oscillation periods. Final data is shown for VI and RK2, for 
low and high resolutions.
}
\label{phasespace}
\end{figure*}

For our next example, consider the simple pendulum with Hamiltonian
\begin{equation}
  H = \frac{1}{2} p^2 - \cos(x) \ ,
\end{equation}
where $x$ denotes the angle from the vertical and $p$ is the angular momentum. 
Figure \ref{phasespace} shows a portion of the phase space 
for the system. We consider a family of initial data points clustered about 
$x = \pi/2$, $p = 0$. Specifically, the initial data are given by 
\begin{equation}
  x = \pi/2 + 0.002\cos(\theta) \ ,\qquad p = 0.002\sin(\theta)
\end{equation}
for $0 \leq \theta \leq 2\pi$. These points form a ``circle'' in 
phase space. The boxes in Fig.~\ref{phasespace} mark the points 
$\theta = 0$, $\pi/2$, $\pi$, and $3\pi/2$. 
The initial data are evolved with 
VI  and with RK2, both at low resolution (timestep $\Delta t = 0.1$) and 
high resolution (timestep $\Delta t = 0.05$). The run time is $74.1$ time units, which 
is just under $10$ oscillation periods. The initial data cycles around the phase 
space diagram in a clockwise direction. Figure \ref{phasespace} shows the end result of 
this evolution for the two integrators at low and high resolutions, as well as the 
``exact'' solution obtained from RK4 with a very small timestep. The dashed 
curves show the constant energy contours with energies determined  by the initial 
data shown as boxes. 

Qualitatively, we see that both VI and RK2 schemes are second order accurate. That is, 
the errors in $x$ and $p$ are reduced by a factor of about $4$ when the resolution is doubled. But 
the character of that error is very different. The VI evolution stays close to the constant energy contours, 
and the phase space errors lie almost entirely in the $H = {\rm constant}$ subspace. 
The RK2 integrator does not respect conservation of energy, and over time the system point 
in phase space spirals outward with increasing energy.  
After about $9000$ time units the simulation 
with  RK2 and $\Delta t = 0.1$ predicts that the pendulum will gain enough 
energy to circle around completely, rather than oscillate. 

Recall that the midpoint rule is a symplectic integrator, that is, the symplectic form 
(\ref{VIsymplecticform}) is preserved
in time. It follows that the volume of phase space bounded by  the initial data ``circle'' 
in Fig.~\ref{phasespace} is constant under the discrete evolution defined by the variational integrator. 
The standard second order Runge--Kutta scheme is not symplectic, and does not preserve 
phase space volume. In Fig.~\ref{phasespace} it is not possible to tell, simply by looking, whether or 
not the initial phase space volume is conserved by the VI scheme, or changed by the RK2 scheme. 
A more involved numerical test would be needed to verify the expected results. 

\subsection{Unbounded motion in one dimension}
The VI equations conserve the phase space function ${\cal H}$ exactly, but the energy 
$H$ might not remain close to ${\cal H}$   if the motion of the system is
unbounded. Consider the Hamiltonian for a particle moving in a one dimensional potential, 
$H = p^2/2 + V(x)$. In this case Eq.~(\ref{calHfunctionofH}) gives 
\begin{equation}\label{Hdiffexample}
  H - {\cal H}  = \frac{\Delta t^2}{24} \left[ p^2 V'' + (V')^2 \right] + {\cal O}(\Delta t^4) \ ,
\end{equation}
where prime denotes $d/dx$. 
The time derivative of this difference is $d(H-{\cal H})/dt = p^3 V''' \Delta t^2/24$ plus terms 
of higher order in $\Delta t$. 
We see that $H - {\cal H}$, and therefore also $H$, will grow in time if $p^3 V'''$ 
remains finite and does not change sign. 

A nice example of this unbounded behavior is obtained with the potential $V(x) = -x^{6/5}$. In this 
case the particle motion at late times 
is given approximately by $x \sim (2\sqrt{2} t/5)^{5/2}$, $p \sim \sqrt{2}(2\sqrt{2} t/5)^{3/2}$.  
Equation (\ref{Hdiffexample}) shows that the energy grows linearly with 
time, $H \sim {\cal H} + (2\sqrt{2}\Delta t^2/125) t$. 
\begin{figure}[!htb]
\begingroup%
  \makeatletter%
  \newcommand{\GNUPLOTspecial}{%
    \@sanitize\catcode`\%=14\relax\special}%
  \setlength{\unitlength}{0.1bp}%
\begin{picture}(2448,1943)(0,0)%
\special{psfile=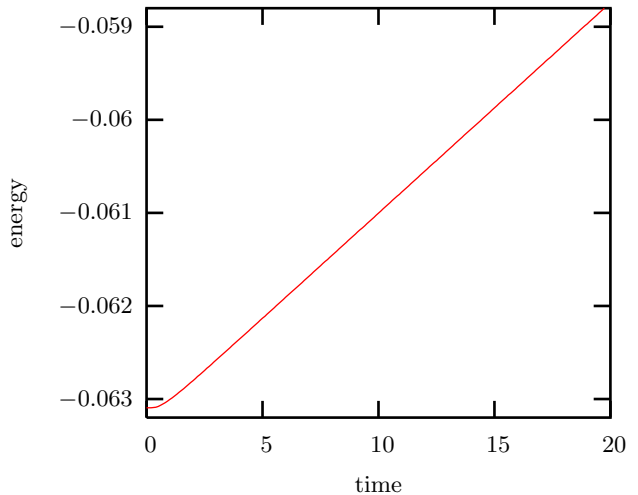 llx=0 lly=0 urx=245 ury=194 rwi=2450}
\put(1424,50){\makebox(0,0){time}}%
\put(100,1072){%
\special{ps: gsave currentpoint currentpoint translate
270 rotate neg exch neg exch translate}%
\makebox(0,0)[b]{\shortstack{energy}}%
\special{ps: currentpoint grestore moveto}%
}%
\put(2298,200){\makebox(0,0){ $20$}}%
\put(1861,200){\makebox(0,0){ $15$}}%
\put(1424,200){\makebox(0,0){ $10$}}%
\put(987,200){\makebox(0,0){ $5$}}%
\put(550,200){\makebox(0,0){ $0$}}%
\put(500,1774){\makebox(0,0)[r]{$-0.059$}}%
\put(500,1423){\makebox(0,0)[r]{$-0.06$}}%
\put(500,1072){\makebox(0,0)[r]{$-0.061$}}%
\put(500,721){\makebox(0,0)[r]{$-0.062$}}%
\put(500,370){\makebox(0,0)[r]{$-0.063$}}%
\end{picture}%
\endgroup
 
\caption{Energy as a function of time for a particle in a one--dimensional 
potential, $V(x) = -x^{6/5}$, obtained with VI. 
As expected, the error in energy grows linearly with time.}
\label{linHerr}
\end{figure}
Figure \ref{linHerr} confirms that for this system, the variational integrator 
exhibits linear growth in the energy.  The initial data used in this
simulation was $x = 0.1$, $p = 0$, with timestep $\Delta t = 0.1$. 
The energy error obtained from RK2 is almost identical to the result shown 
in Fig.~\ref{linHerr} for VI. 

\subsection{Orbital motion}
Our final example is motion in a gravitational (or electric) 
field described by a central $1/r$ potential. The 
Hamiltonian is defined by 
\begin{equation}
  H = \frac{1}{2}(p_x^2 + p_y^2) - \frac{1}{\sqrt{x^2 + y^2}}
\end{equation}
This system is symmetric under rotations in the $x$--$y$ plane. 
The conserved Noether charge associated with rotational symmetry is  
angular momentum, $J \equiv x p_y - y p_x$. The initial data 
for this simulation is $x = 1.0$, $p_x = 0.0$, $y = 0.0$, 
$p_y = 1.2$. The resulting orbital motion is an ellipse with 
eccentricity $\sim 0.5$ and period $\sim 15$. 

\begin{figure}[!htb]
\begingroup%
  \makeatletter%
  \newcommand{\GNUPLOTspecial}{%
    \@sanitize\catcode`\%=14\relax\special}%
  \setlength{\unitlength}{0.1bp}%
\begin{picture}(2448,1943)(0,0)%
\special{psfile=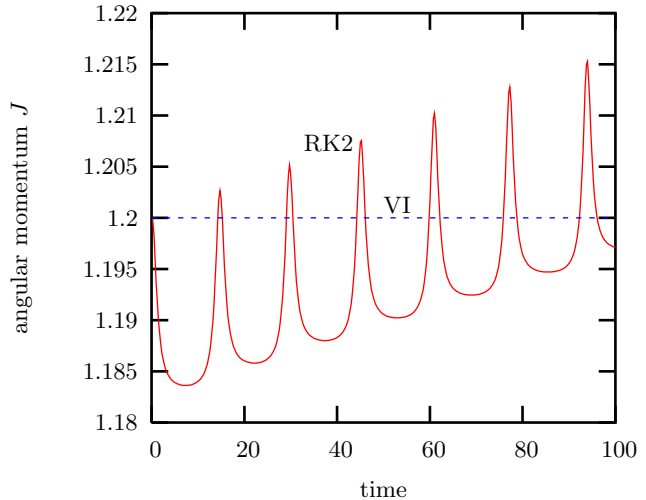 llx=0 lly=0 urx=245 ury=194 rwi=2450}
\put(1425,1120){\makebox(0,0)[l]{VI}}%
\put(1125,1355){\makebox(0,0)[l]{RK2}}%
\put(1424,50){\makebox(0,0){time}}%
\put(100,1072){%
\special{ps: gsave currentpoint currentpoint translate
270 rotate neg exch neg exch translate}%
\makebox(0,0)[b]{\shortstack{angular momentum $J$}}%
\special{ps: currentpoint grestore moveto}%
}%
\put(2298,200){\makebox(0,0){ $100$}}%
\put(1948,200){\makebox(0,0){ $80$}}%
\put(1599,200){\makebox(0,0){ $60$}}%
\put(1249,200){\makebox(0,0){ $40$}}%
\put(900,200){\makebox(0,0){ $20$}}%
\put(550,200){\makebox(0,0){ $0$}}%
\put(500,1844){\makebox(0,0)[r]{ $1.22$}}%
\put(500,1651){\makebox(0,0)[r]{ $1.215$}}%
\put(500,1458){\makebox(0,0)[r]{ $1.21$}}%
\put(500,1265){\makebox(0,0)[r]{ $1.205$}}%
\put(500,1072){\makebox(0,0)[r]{ $1.2$}}%
\put(500,879){\makebox(0,0)[r]{ $1.195$}}%
\put(500,686){\makebox(0,0)[r]{ $1.19$}}%
\put(500,493){\makebox(0,0)[r]{ $1.185$}}%
\put(500,300){\makebox(0,0)[r]{ $1.18$}}%
\end{picture}%
\endgroup
 
\caption{Angular momentum as a function of time for motion in a 
central potential. The solid curve is obtained from RK2. The 
constant, dashed line is obtained with the variational integrator.}
\label{angmom}
\end{figure}

Figure \ref{angmom} 
shows the angular momentum as a function of time for RK2 and VI 
with timestep $\Delta t = 0.25$. 
With the variational integrator the angular momentum is exactly 
conserved (to machine accuracy) and $J$ retains its initial value of $1.2$ throughout 
the simulation. With RK2, the angular momentum exhibits short timescale 
fluctuations and a longer timescale drift. A more complete analysis 
shows that the short timescale errors are order $\Delta t^2$, whereas 
the drift in $J$ is order $\Delta t^3$. Qualitatively similar results 
are obtained for RK4. In that case, the short timescale errors in $J$ are order 
$\Delta t^4$, and the drift is order $\Delta t^5$. 

\begin{acknowledgments}
This work was supported by NASA Space Sciences Grant ATP02--0043--0056. 
\end{acknowledgments}

\bibliography{references}

\end{document}